\documentclass[prl,twocolumn,showpacs,superscriptaddress]{revtex4}
\usepackage{graphicx}
\usepackage{dcolumn}

\begin{document}
\title{Periodical Plasma Structures Controlled by Oblique 
 Magnetic Field}
\author{I. V. Schweigert}
\affiliation{George Washington University, Washington D.C. 
20052, USA}
\affiliation{Khristianovich Institute of Theoretical and 
Applied Mechanics, Novosibirsk 630090, Russia}
%\author{J. Lukas}
\author{M. Keidar}
\affiliation{George Washington University, Washington D.C. 
20052, USA}

\date{\today }            

\begin{abstract}
 The propulsion type plasma in oblique external magnetic field 
 is studied in 2D3V PIC MCC simulations. 
 A periodical structure  with maxima of electron and ion densities
 appears with an increase of an obliqueness 
 of magnetic field. These 
 ridges of electron and ion densities are aligned with 
 the magnetic field vector and shifted relative each other.
 As a result the two-dimensional double-layers 
 structure forms in cylindrical plasma chamber.  
 The ion and electron currents on the side wall are essential modulated 
 by the oblique magnetic field.

\end{abstract}

\maketitle

%%%%%%%%%%%%%%%%%%%%%%%%%%%%%%%%%%%%%%
\section{Introduction}
%%%%%%%%%%%%%%%%%%%%%%%%%%%%%%%%%%%%%%
 Recently some methods to control the Hall effect thruster
 characteristics with applying the oblique magnetic field with 
 respect to the channel walls  is widely discussed 
 (see, \cite{Mied,Walker} and references cited 
 therein).
 Nevertheless with increasing the inclination of the magnetic field, 
 discharge plasma properties can essentially change.
 For example, a several stationary, magnetized, two-dimensional 
 weak double-layers  were observed in a laboratory experiment 
 for this type of plasma  by Intrator, Menard, Hershkowitz 
 \cite{Intrator}. 
 The  double-layer potential drops were found to be followed across 
 any given potential profile. Borovsky, Joyce \cite{Borovsky} showed
in PIC simulations that weak magnetization results in the double-layer
electric-field alignment of particles accelerated 
by these potential structures
and that strong magnetization results in their magnetic-field alignment.
 A morphological invariance in two-dimensional
double-layers with respect to the degree of magnetization observed 
 in Ref. \cite{Borovsky} implied that the
potential structures scale with Debye lengths rather than with gyroradii.

A weak double-layer is a nonlinear electrostatic structure in
plasmas, consisting of two sheets of positive and negative charges, with a
characteristic electric potential jump, providing local electric
field. Recently, most of the studies have addressed strong or ion
acoustic double-layer in  magnetized plasmas
(see for example \cite{2,3,4,9,10},  while the mechanism of 
 weak two dimensional double-layer formation is still not fully 
understood.

The motivation of our recent study is irregularities observed in 
Refs. \cite{Intrator,disser} in bounded low pressure discharge plasma induced by 
application of an oblique  magnetic field.  
These irregularities seems similar  to  aurora-like periodical forms 
induced in ionosphere plasma with solar wind and interplanetary 
magnetic field (see \cite{aurora} and references therein). 

 In this paper, in kinetic simulations  we consider 
the direct current discharge plasma in the external oblique
magnetic field at low gas pressure, P=10$^{-4}$ Torr.  
 Our purpose is to study the plasma structure modification with 
 changing the obliqueness and strength of  
 magnetic field for the plasma parameters
 similar to the Hall thruster ones.
 We also consider the effect of obliqueness of magnetic field 
of the plasma-wall transition sheath in Hall thruster type plasma 
 since the magnetic lens can be used to focus the ion flux. 

\section{Theoretical Model and Calculation  Details} 

\begin{figure}[h!]
\includegraphics[width=0.75\linewidth]{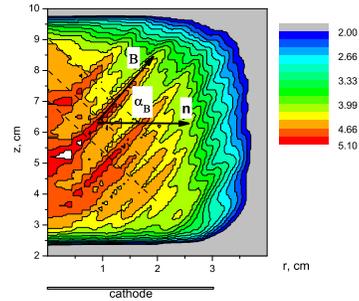}
\caption{Distribution of potential (V)
  for B=50~G, $\alpha_B$ = 65$^{\circ}$ and  $T_e$=5~eV.
  B is the magnetic field vector, $n$ is a normal to the side wall,
 $\alpha_B$ is the angle between $B$-vector and $n$.
 Cathode is at z = 0.3 cm.}
\label{setup}
\end{figure}

 In our simulations, the plasma is embedded in a cylindrical 
 chamber with the radius of 4~cm  and the height of 10 cm. 
 The calculation domain is shown  in Fig.\ref{setup}.
 The  cathode with the radius of 3~cm is placed  0.3~cm apart from
 the chamber bottom. 
 All walls of the chamber are grounded and the cathode is biased
 with -90 V.   The strength of external magnetic field B is assumed 
 to be constant  over the plasma volume. The magnetic field is
 axially symmetrical. To avoid the singularity 
 at r=0, we took the
 magnetic field angle  $\alpha_B$ in the following form:
 $\alpha_B$=0 at r$<$0.3~cm, 
 $\alpha_B=\alpha_{B0}$  at r$>$0.6~cm,
 $\alpha_B$  is approximated by a spline-function at 
 0.3~cm$<$r$<$0.6~cm. 
 The $\alpha_{B0}$ ranges from 0 to 77$^{\circ}$ in different variants.

  To describe  the plasma in electro-magnetic field at low gas pressure
  we solve Boltzmann equations (two dimensional in space and three 
 dimensional in velocity space) for the  distribution functions for  electrons   
$f_e(\vec r,\vec v)$ 
and ions $f_i(\vec r,\vec v)$  
\begin{equation}  \label{kine}
\frac {\partial f_e}{\partial t}+ \vec v_e\frac {\partial 
f_e}{\partial \vec r}
-\frac {e(\vec E+\vec v_e\times\vec B)}{m}\frac {\partial f_e}{\partial \vec 
v_e}= 
J_e,\quad n_e=\int
f_ed\vec v_e,
\end{equation}
\begin{equation}  \label{kini}
\frac {\partial f_i}{\partial t}+ \vec v_i\frac {\partial 
f_i}{\partial \vec r}
+ \frac {e(\vec E+\vec v_i\times\vec B)}{M}\frac {\partial f_i}{\partial \vec 
v_i} 
=J_i,\quad
n_i=\int f_id\vec v_i,
\end{equation}
where $v_e$, $v_i$, $n_e$, $n_i$, $m$, $M$ are the electron and ion velocities,
 densities and masses, respectively.
$J_e$, $J_i$ are the collisional integrals for electrons and ions.
  No magnetic field due to currents 
 in the plasma is taken into account. 
The Poisson equation describes the electrical potential and
 electrical field distributions
\begin{equation}  \label{Poisson}
\bigtriangleup \phi =4\pi e \left(n_e-n_i \right),
\quad
\vec E=-\frac{\partial \phi}{\partial \vec r} \; .
\end{equation}

  The system of equations (1)-(3) is
  solved with the 2D3V Particle-in-cell Monte Carlo collision method 
 (PIC MCC) with PlasmaNOV code \cite {SCH}. For electrons the elastic 
  scattering, excitation and ionization were taken into account with 
 the cross sections from Refs.\cite{Ivan,Lagush}. For ions the resonant charge
 exchange collisions  with background argon gas are included.
 First, solving Eqs. (1)-(2), we calculate the electrical charge 
 distribution, then  the Poisson equation is solved to find a map 
 of the electrical potential. The Poisson equation is solved on each electron 
 time step.
 The steady-state solution is reached by iteration method.

 The boundary conditions are the following: 
 $\phi$ =-90~V at the cathode,  $\phi$ =0 at the wall of chamber and
 $\delta\phi/\delta r$=0 at r=0. 
 Ions and electrons approaching the surface disappear from the calculation 
 domain. The electron emission from the surface is not included 
 in the model.

 Additionally to the electron impact ionization the external ionization 
 is included in the model.
 The external ionization is modeled as electron-ion pairs generation with
 the Maxwellian distributions over velocity with the mean 
 electron temperature $T_e$
 and the ion temperature $T_i$ = 0.05 eV. 
 The electron temperature $T_e$  varies from
 2.5 eV to 10 eV for different cases.
 The rate $\nu_i$ of the electron-ion pair generation is chosen 
  to achieve the plasma density of 
  $10^8$cm$^{-3}$ in quasineutral part.
 For all cases the rate $\nu_i$ is equal to 
 2.5$\times$10$^8$s$^{-1}$cm$^{-3}$.

 In simulations, a strength of magnetic field $B$ is ranged from 
 25~G to 100~G and the angle $\alpha_B$ = 0 - 77$^{\circ}$.    
 For these plasma parameters  the electron Larmor radius $r_L$ is
 comparable to the Debye length $\lambda_D$,
 $r_L \approx \lambda_D$.
 The plasma frequency $\omega_p$ is about of 
 the electron gyrofrequency $\Omega_e$,
 $\omega_p \leq \Omega_e=5\times10^8s^{-1} - 5\times10^9s^{-1}$.
 In simulations, the electron time step $\Delta t_e$ 
 is (2-5)$\times 10^{-12} s$, 
 so $\Delta t_e \ll 1/\omega_p, 1/\Omega_e, \Delta r/v_e, \Delta z/v_e,$
 where $\Delta r$, $\Delta z$ are steps of calculation grid over axes r and z,
 and $v_e$ is the maximum electron velocity.

 The calculation grid is uniform over z-direction and nonuniform 
over radius condensing 
 with increasing r.
 The total number of pseudo particles being chosen 
 so that there is an average of approximately 100 
 positive and negative particles per Debye sphere.
 The electron/ion mass ratio is chosen to be a real one.

\section{Effect of Magnetic Field Angle}
\begin{figure}[h!]
\includegraphics[width=1.\linewidth]{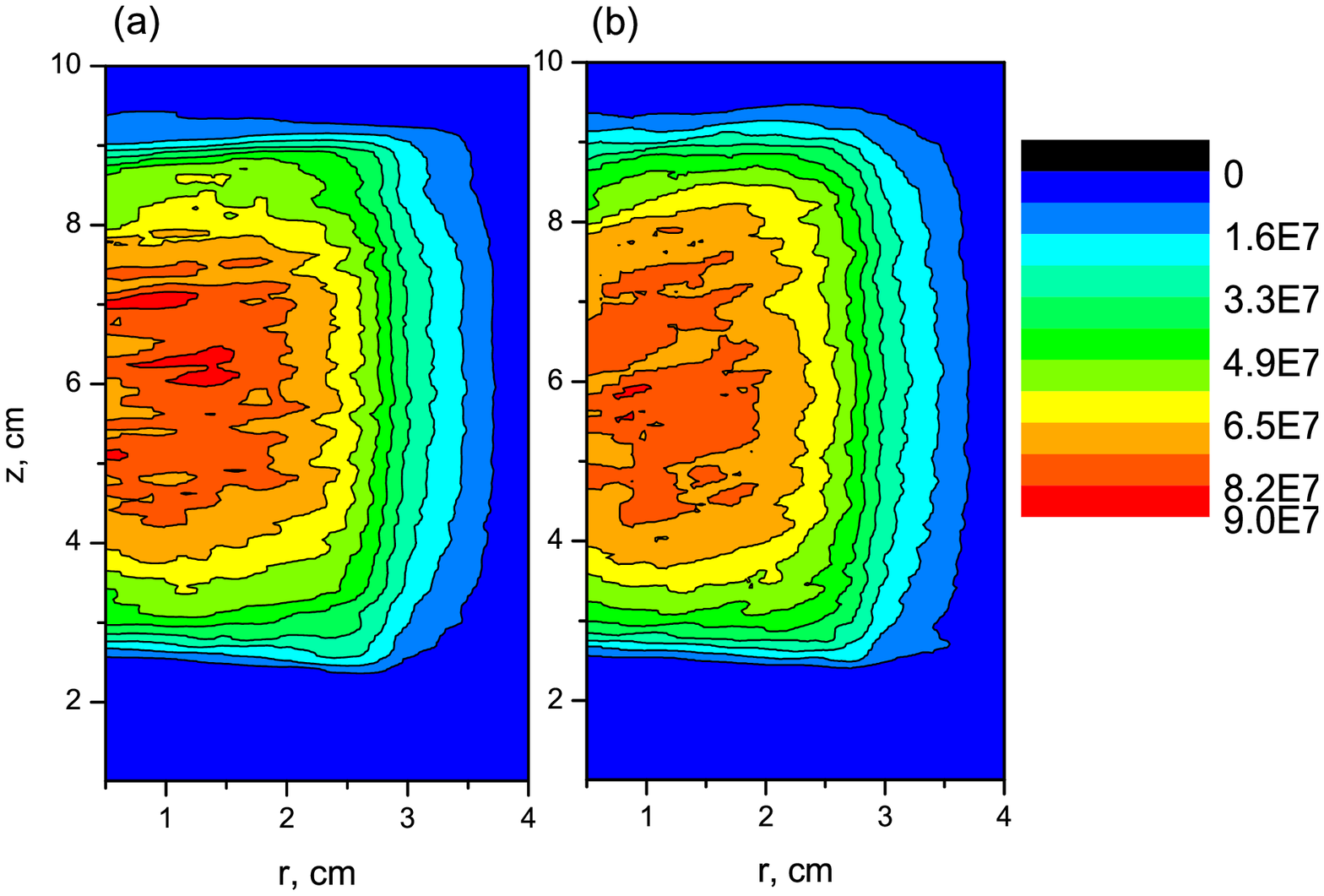}
\includegraphics[width=1.\linewidth]{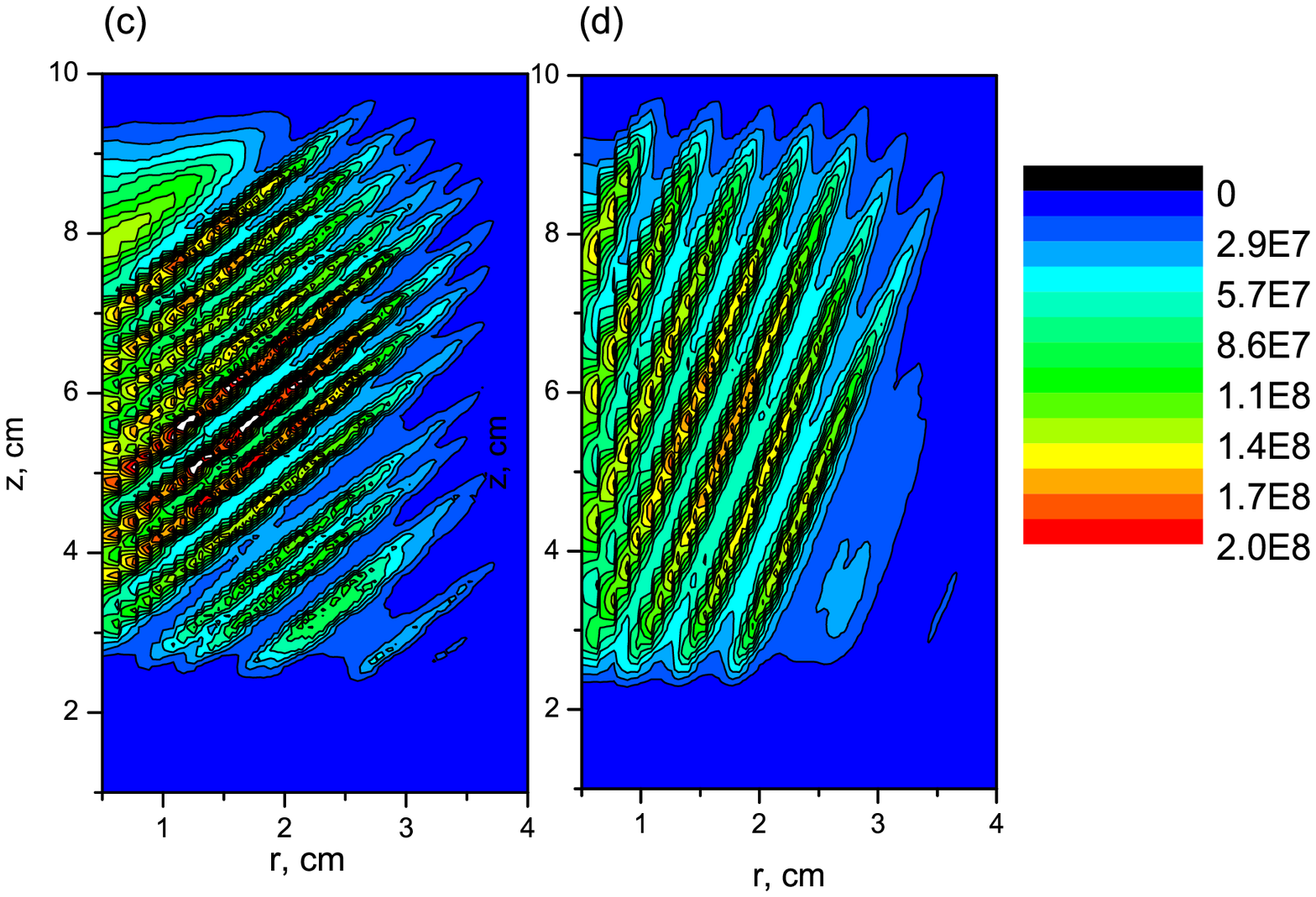}
\caption{Electron density distribution $n_e$, cm$^{-3}$,
  for $\alpha_B$= 10$^{\circ}$(a), 27$^{\circ}$(b), 
  55$^{\circ}$(c) and 77$^{\circ}$(d),  
 B=50 G, $T_e$=2.5 eV.}
\label{angle}
\end{figure}

 In PIC MCC simulations, we found that a rearrangement of plasma begins
  with increasing obliqueness of B-field. 
 For small $\alpha_B$=10$^{\circ}$ and 27$^{\circ}$ 
 shown in Fig.\ref{angle}(a) and (b), 
  the electron density is almost uniform in the central part of 
 the chamber. 
 In general, the plasma looks very similar
 to the case of $\alpha_B$=0. 
 The developed sheath forms near the cathode with the potential drop
 of approximately 93 V. A weaker sheath screens plasma from walls 
 of the chamber  with the potential drop $\delta \phi_w\approx$ 3 V.
 The sheaths can be seen in Fig.\ref{angle} as areas with 
 the depleted electron density. 

  With increasing $\alpha_B$ the periodical plasma structure becomes 
 clearly visible  (see Fig.\ref{angle}(c) and (d)).
  The structure occupies the quasineutral part of plasma 
  in which the electrical field is small. 
  Within the cathode sheath it does not appear 
   even for large $\alpha_B$. More complex situation takes place
  in the wall sheath. 
  As the potential drop over the wall is small
  ($\delta \phi_w$=3~V - 8~V), 
  the Lorentz and electrical forces are comparable there.
  Therefore the wall sheath becomes modulated by B-field with increasing
  $\alpha_B$.

\begin{figure}[h!]
\includegraphics[width=0.8\linewidth]{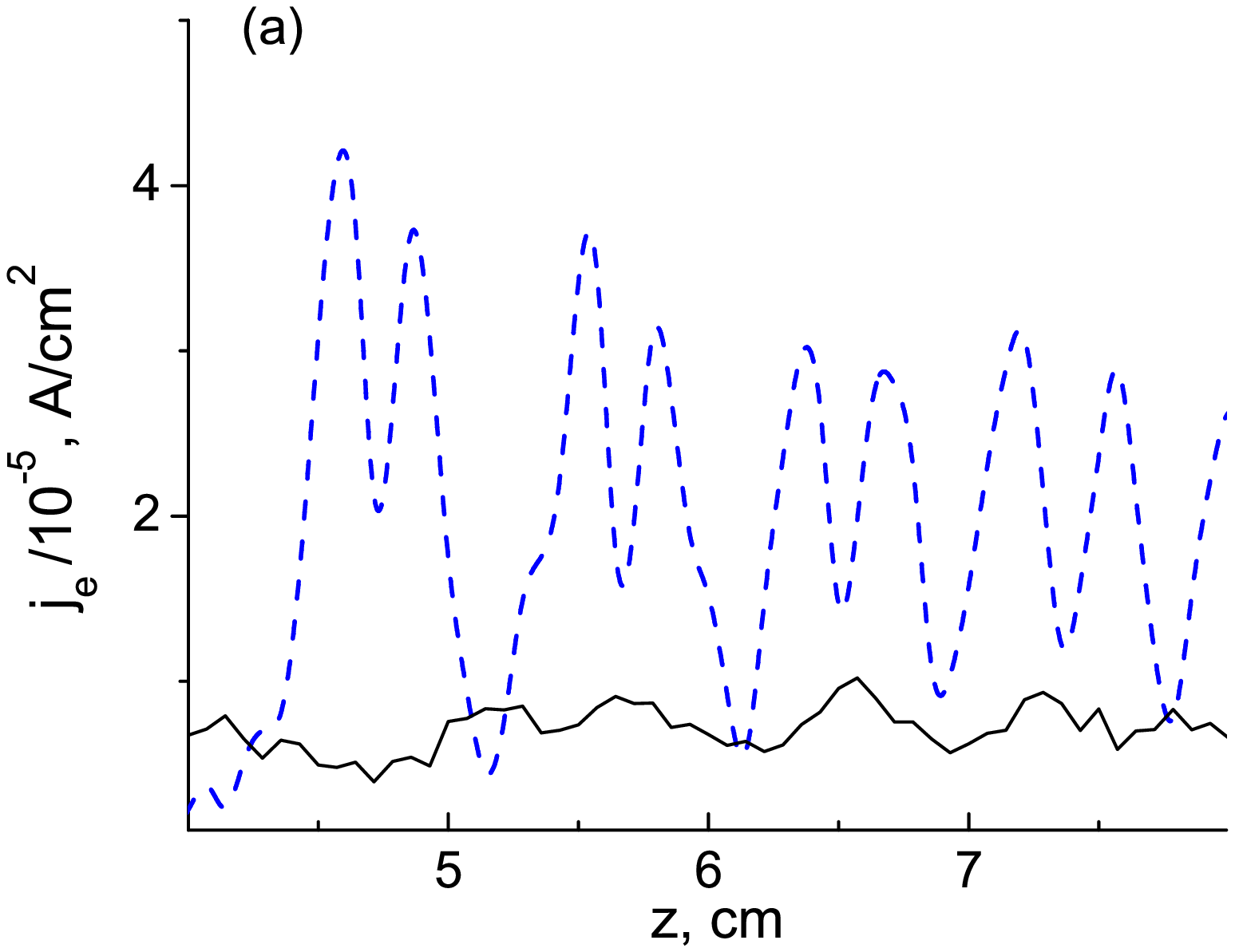}
\includegraphics[width=0.8\linewidth]{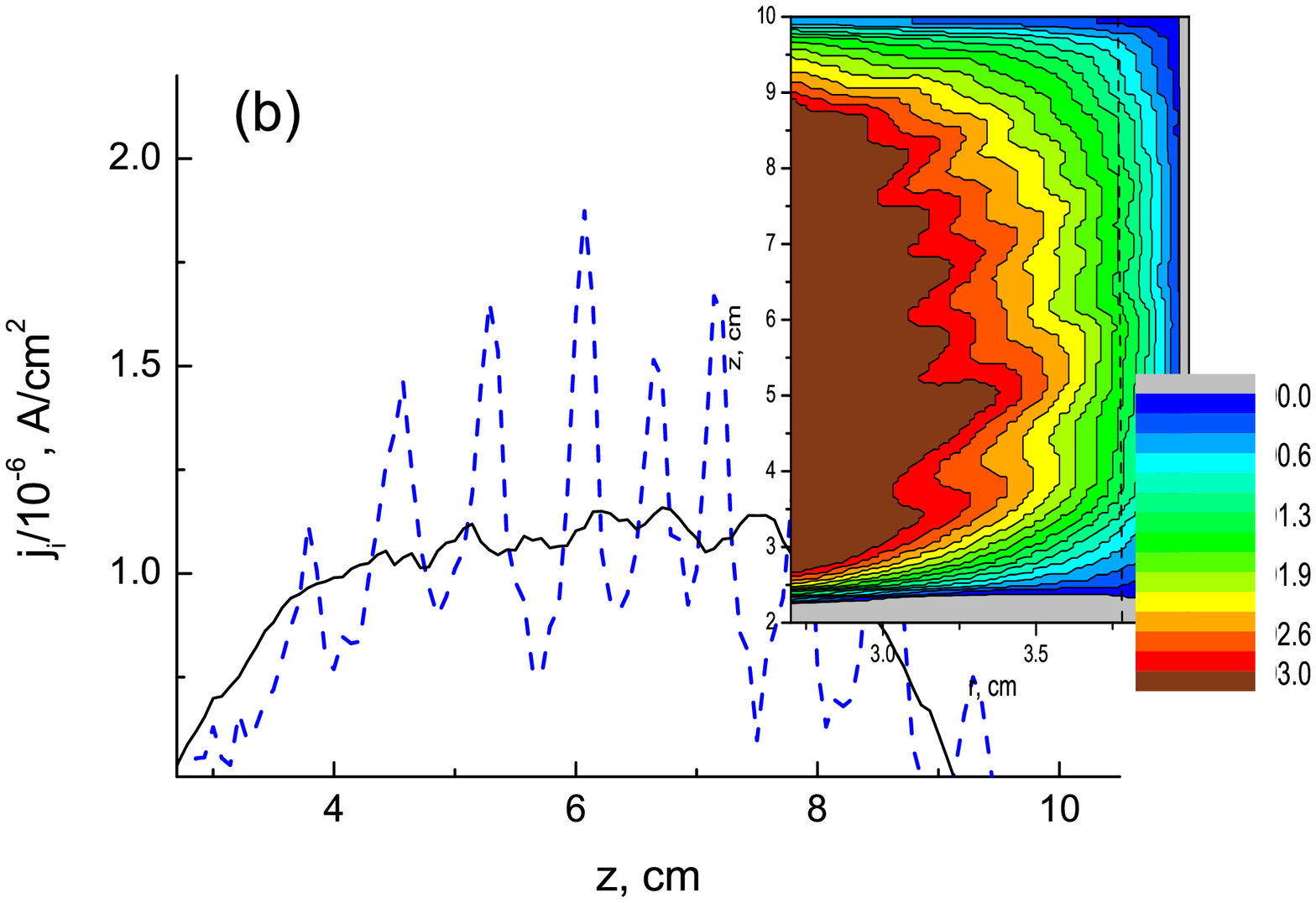}
\includegraphics[width=0.8\linewidth]{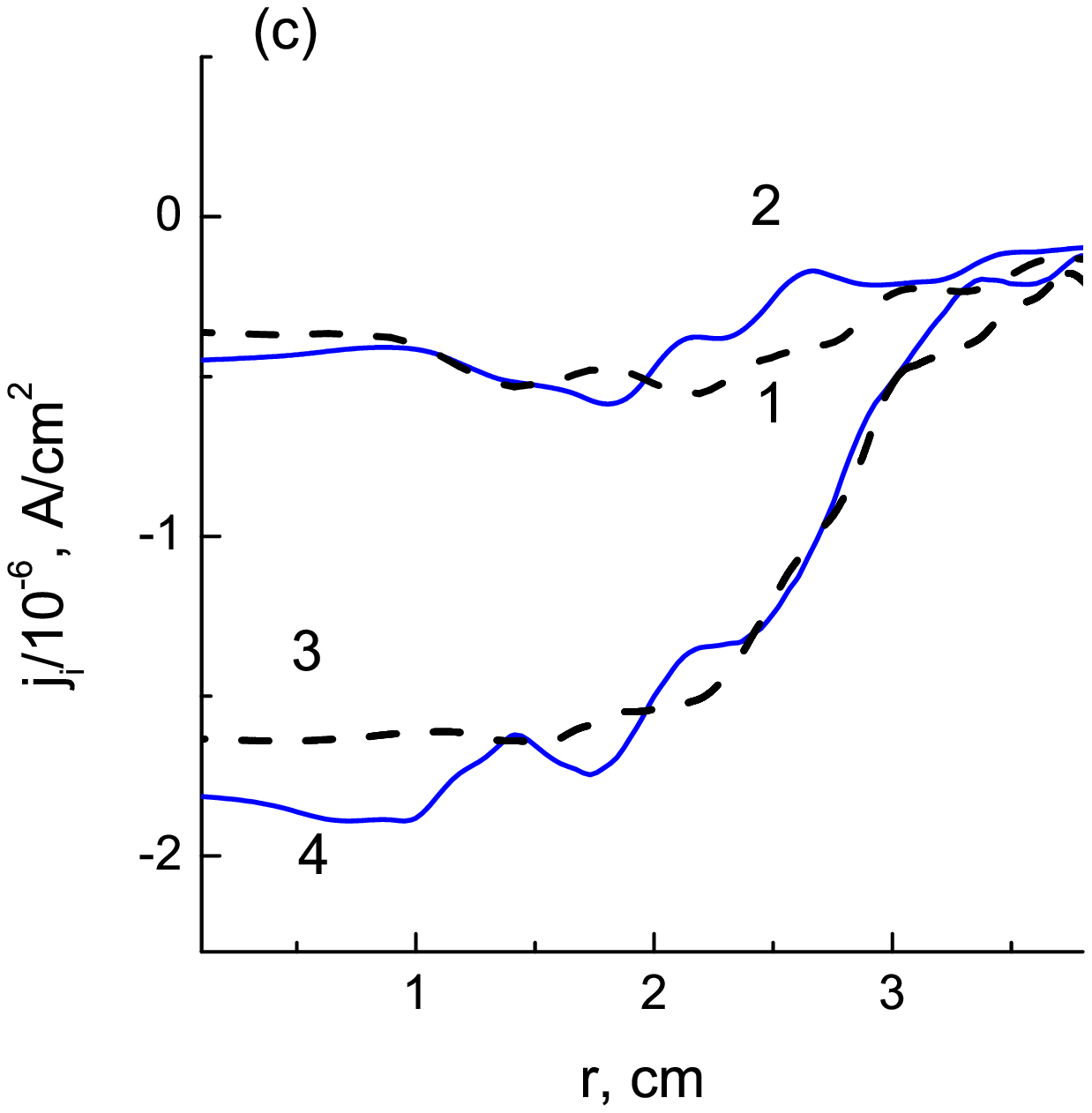}
\caption{(a) Electron current density distribution over z at r=3 cm, 
 (b) ion current density over z near  side wall 
  for $\alpha_B$= 10$^{\circ}$ (solid line) and 
  $\alpha_B$= 65$^{\circ}$ (dashed line),
 insert shows the potential distribution (V) near the side wall
 $\alpha_B$= 65$^{\circ}$,  
 (c) ion current density in the cathode sheath
 at z=3.5 cm (1), 1.5 cm (3) for   $\alpha_B$= 10$^{\circ}$
 and z=3.5 cm (2), 1.5 cm (4) for   $\alpha_B$= 65$^{\circ}$.
}
\label{cur_r}
\end{figure}

  In Fig.\ref{cur_r}(a) and (b), the electron $j_e$ and ion $j_i$ currents near the wall 
  are shown for two values of $\alpha_B$. It is seen that 
  both currents approaching the wall are affected by a variation of $\alpha_B$.
  The $j_e$-profile over z taken at r=3~cm is almost uniform for
   $\alpha_B$=10$^{\circ}$ and has peaks for $\alpha_B$=65$^{\circ}$.
 Each 
 electron current peak is splitted with a scale of 2$r_L$, where 
 $r_L$ is Larmor radius. 
 The $j_i$-profile over z taken near the wall also exhibits peaks for larger
 $\alpha_B$. 
 The $j_i$ is about 20 time less than the $j_e$, 
 but both clearly indicate  the periodical plasma structure. 
 An increase of the ion current and its peaked profile  
 are typically observed in our simulations for larger  $\alpha_B$.
 This effect can lead to an additional local erosion of wall material.

  The insert in Fig.\ref{cur_r}, for  $\alpha_B$=65$^{\circ}$,
shows the potential distribution near the
 side wall. It is seen that the wall sheath is disturbed with
 the oblique magnetic field with large $\alpha_B$. 
  The sheath potential exhibits the nonmonotonic 
 distribution within the presheath which is getting smoother 
 close to the wall.

 The ion current profile within the cathode sheath is given in 
 Fig.\ref{cur_r}(c) for two values of $\alpha_B$=10$^{\circ}$ and 65$^{\circ}$. 
 The ion current increases by factor of 4 from z=3.5~cm to z=1.5~cm,
 both for small and large $\alpha_B$, as 
 the electrical field increases within the sheath 
 closer to the cathode. The some focusing effect of the oblique magnetic field
 on the ion flux can be seen for $\alpha_B$=65$^{\circ}$.

 The current flow and ridges of electron and ion densities  
  are aligned with  $B$-vector not only in the area of quasineutral plasma,
 but also within the sheath over the wall. 
 Only in the cathode sheath the ion current is directed normally to 
 the surface.

\section{Multiple Layers Structure}

\begin{figure}[h!]
%\begin{figure*}   
\includegraphics[width=1.\linewidth] {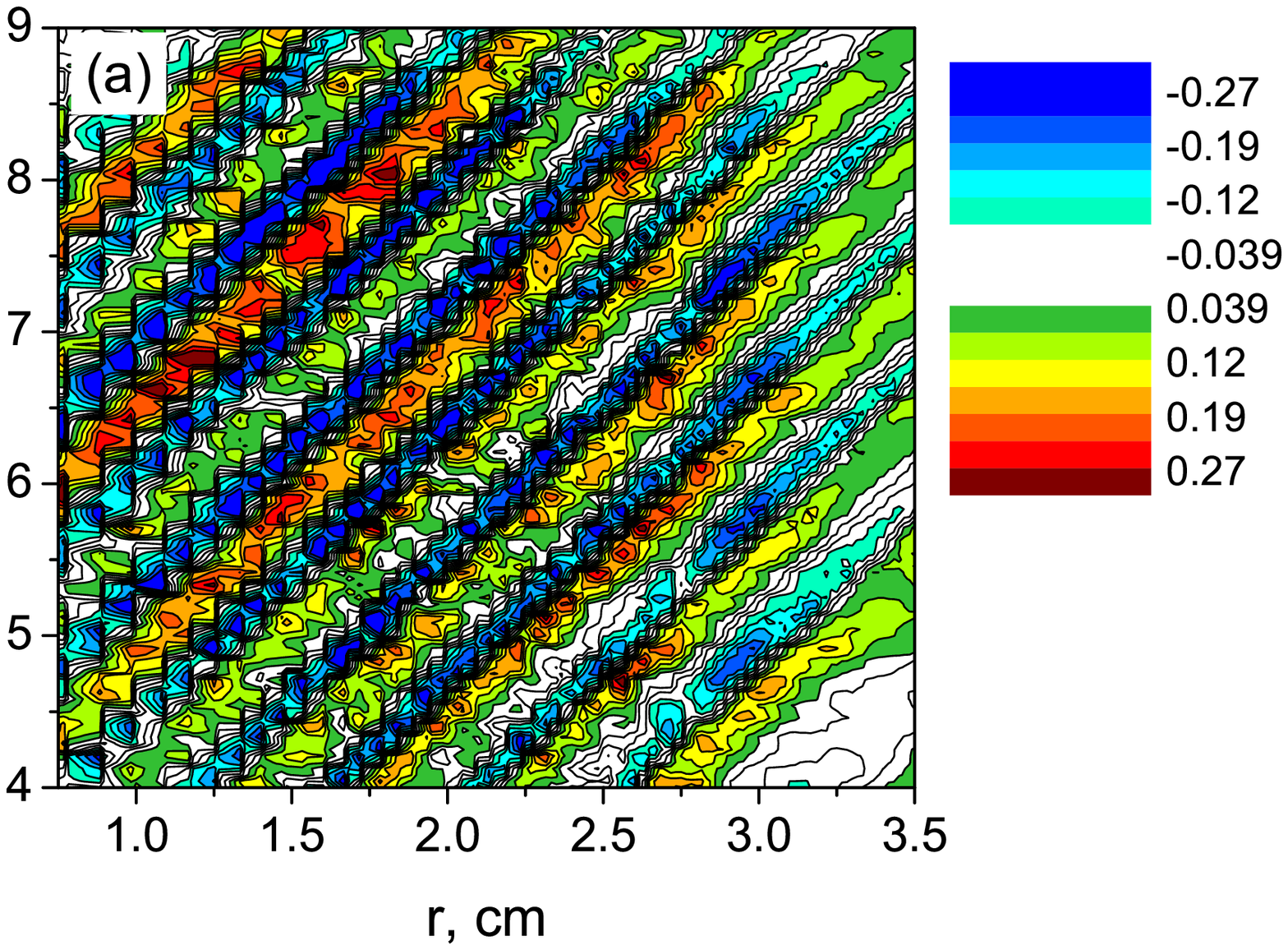}
\includegraphics[width=1.1\linewidth]{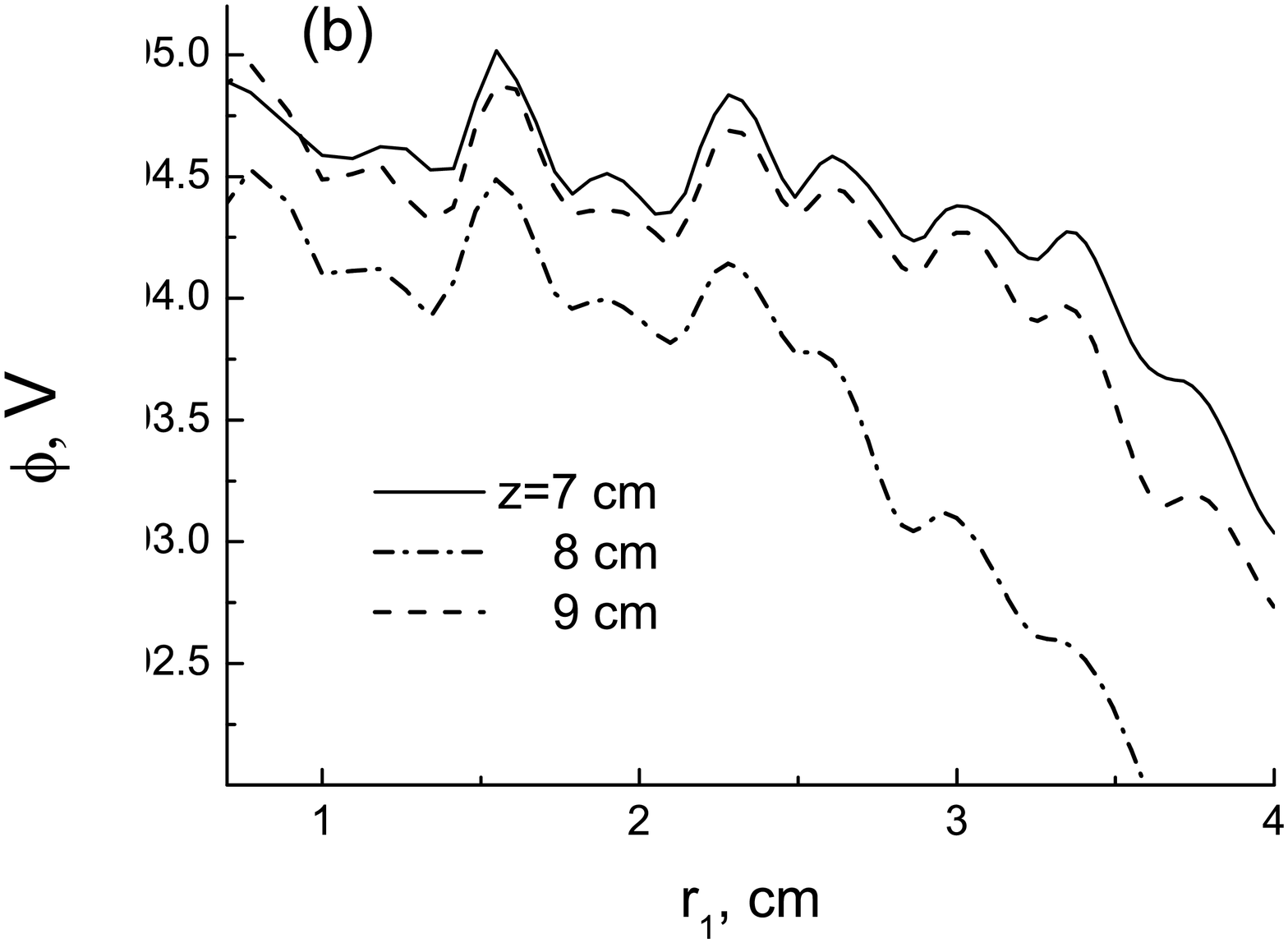}
\includegraphics[width=0.8\linewidth]{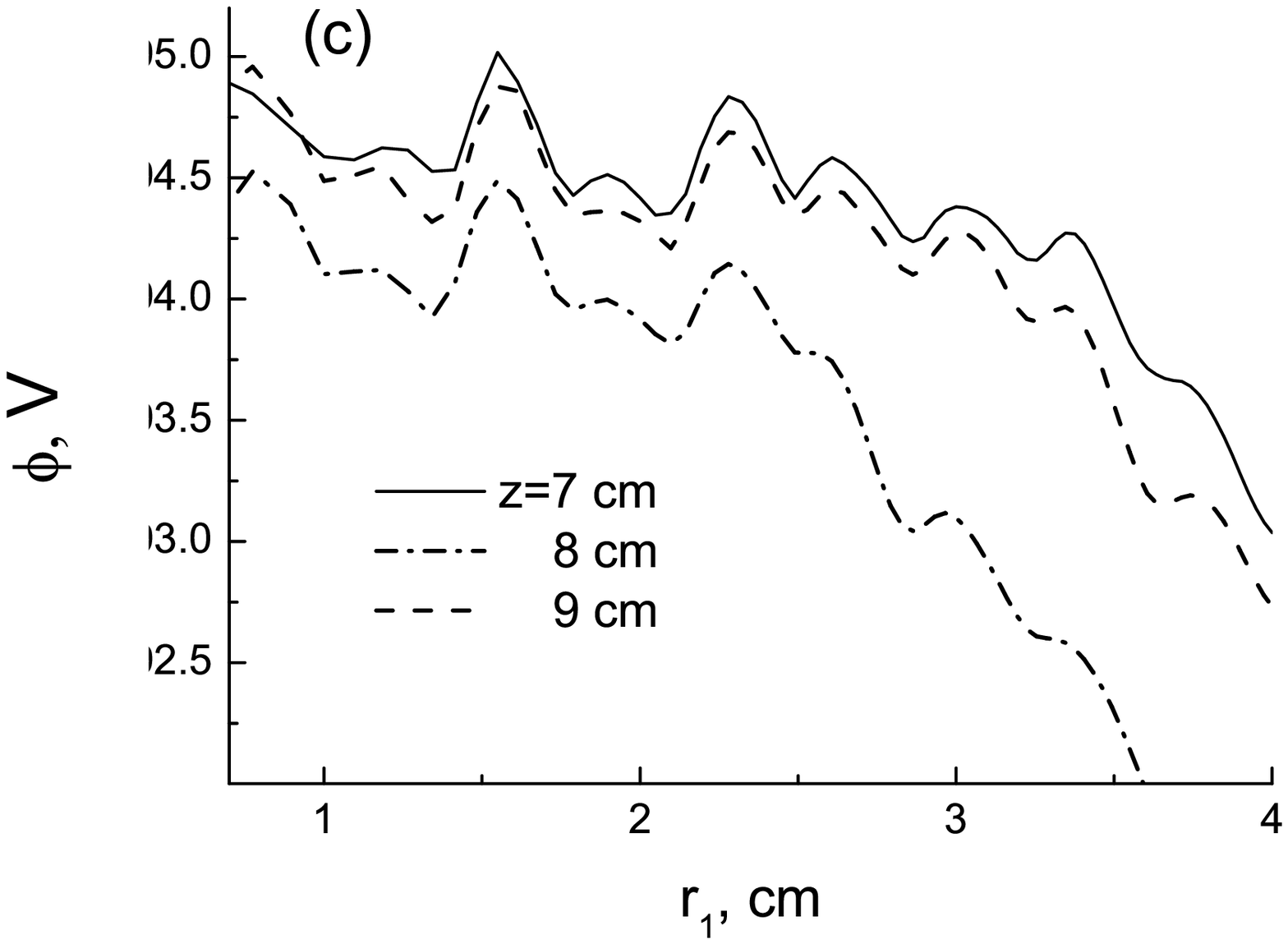}
\caption{
 Distributions of charge ($n_e-n_i$)/10$^8$ cm$^{-3}$ (a), 
  and potential profile along axis $r_p$, 
 which starts from r=0 and 
 z=7 cm (solid line),
 8 cm (dashed line), 9 cm (dashed-dotted line) (b),
 $r_p$ is normal to $B$-vector. 
 B=50 G, $\alpha_B$ = 65$^{\circ}$ and  $T_e$=5 eV.}
\label{pot_q100}
\end{figure}

 In Fig.\ref{pot_q100}, 
 the charge ($n_e-n_i$) and potential distributions 
 are shown for $T_e$=5~eV and B=50~G ($r_L$=0.105~cm).
 The charge distribution  has negative 
 and positive ridges, both with an absolute value of 
 0.27$\times$10$^{8}$cm$^{-3}$. 
 The sheets of large negative and positive charges appear in quasineutral
 plasma due to relative shift of $n_e$ and $n_i$- ridges  
  in the direction of increasing potential and across B-field.
 This structure  is called as double-layers and
 characterized with the non-monotonic potential distribution shown in
 Fig.\ref{setup}. Cross sections of the potential distribution
 across B-field is shown in 
 Fig.\ref{pot_q100}(b). These cross sections are taken along 
 the dashed lines shown in Fig.\ref{setup}, which start at r=0
 and z=7~cm, 8~cm, 9~cm. For this case the potential bumps across  $B$-vector 
 is about 0.5 V. Note that along $B$-vector the potential bumps are smaller
 (0.15 V).  The presence of magnetic field enhances the charge separation 
 across $B$-vector. 

\begin{figure}[h!]
\includegraphics[width=1.\linewidth]{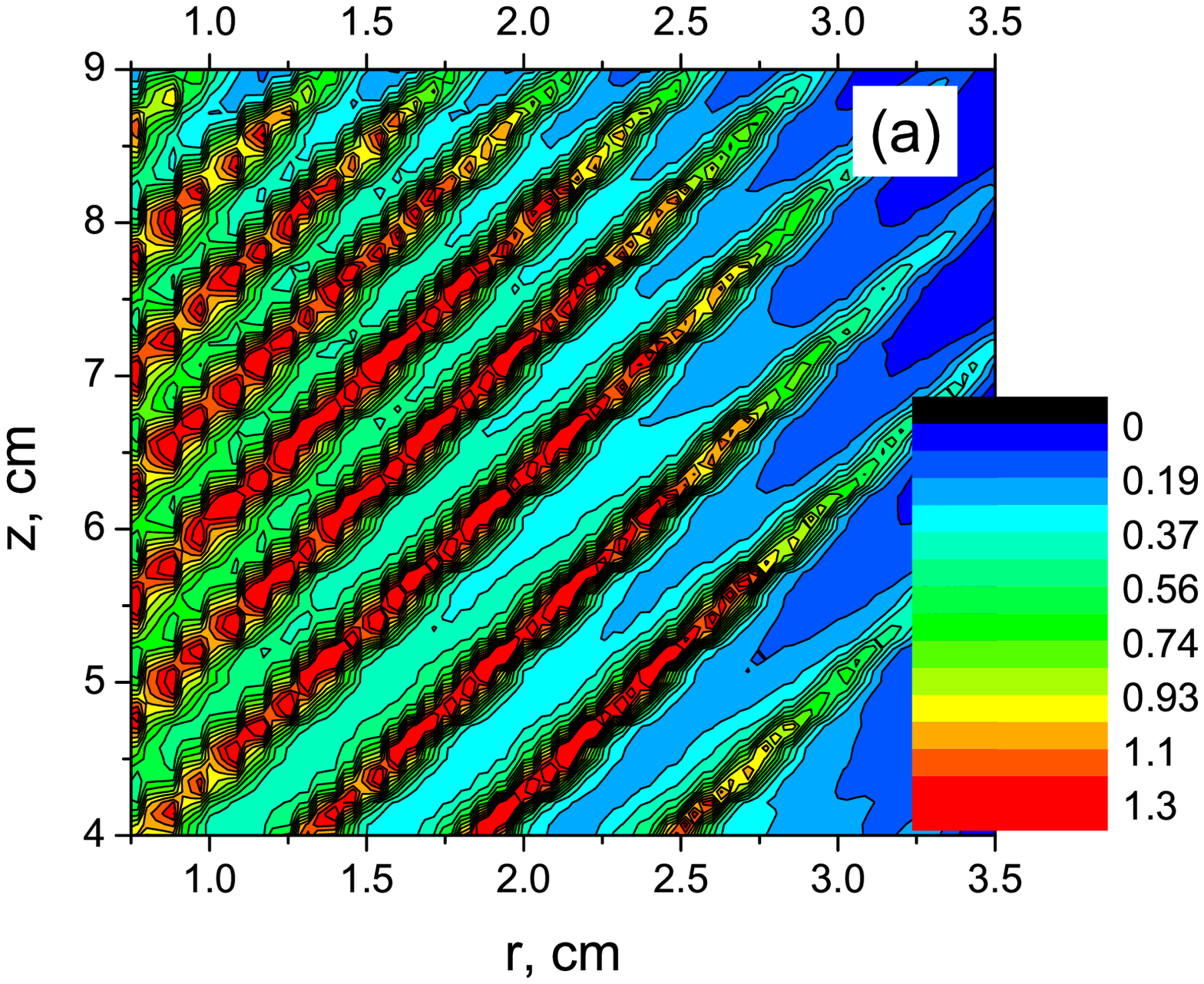}
\includegraphics[width=1.  \linewidth]{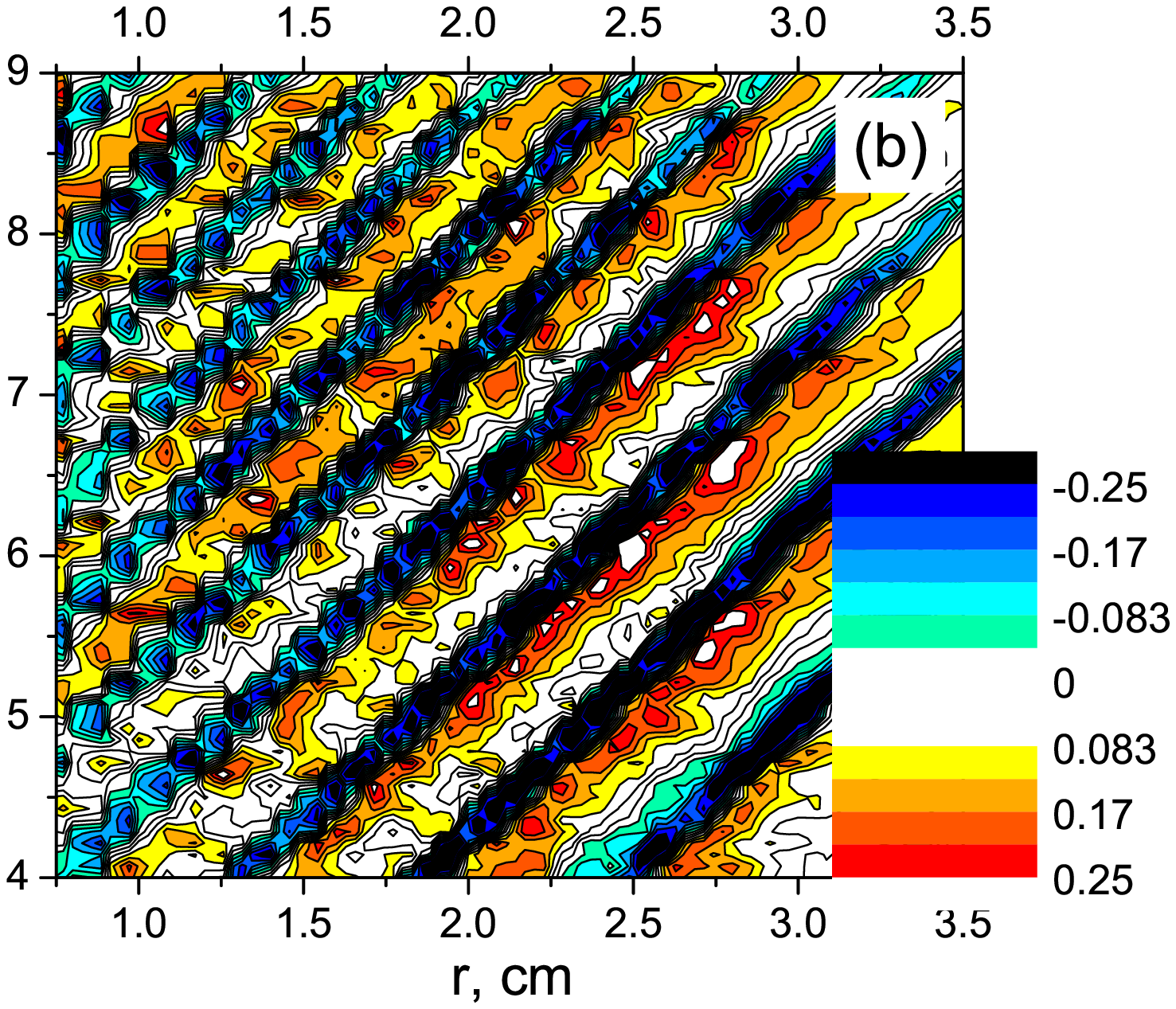}
\includegraphics[width=0.8\linewidth]{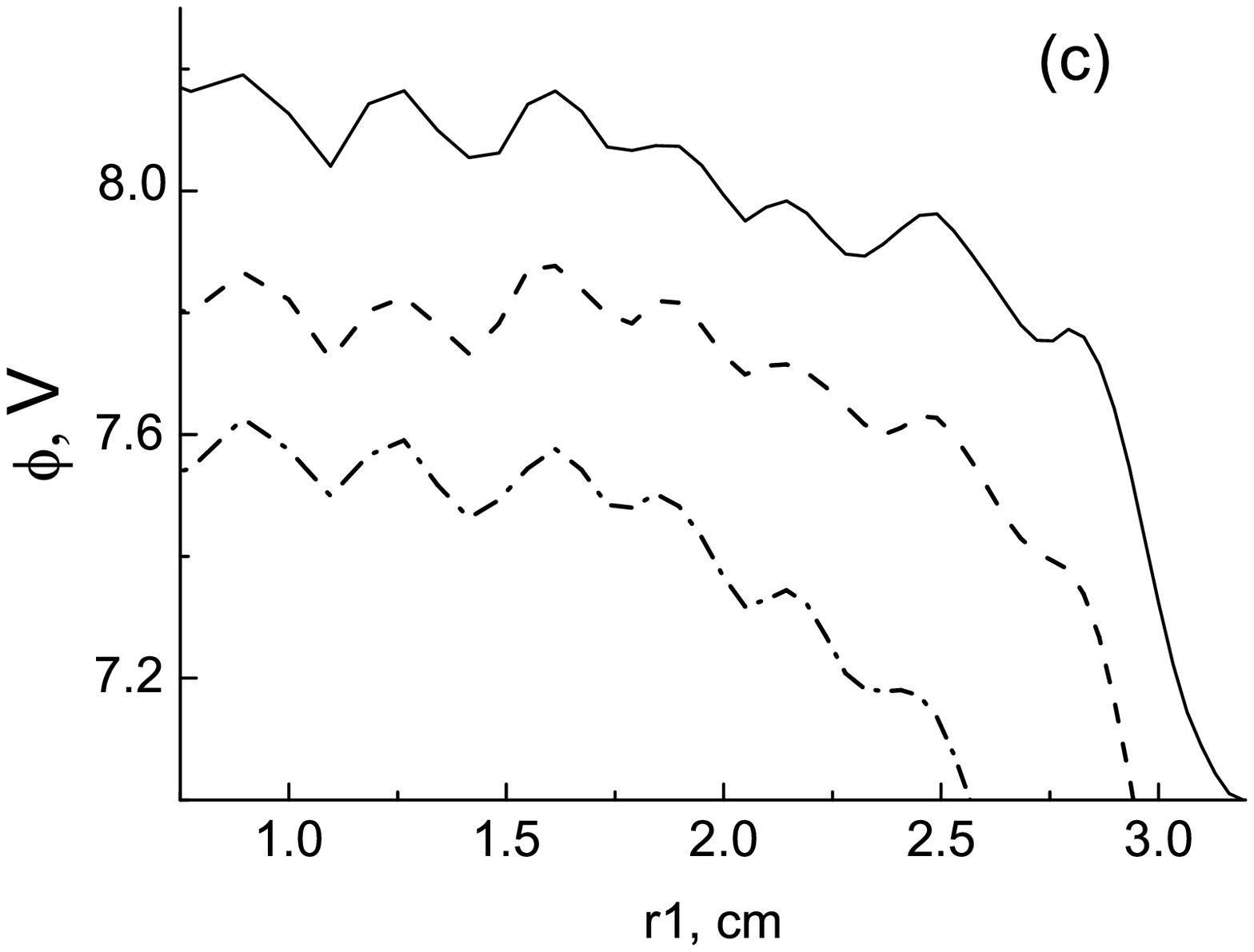}
\caption{
 Distributions of densities of electron $n_e$/10$^8$ cm$^{-3}$ (a) 
 charge ($n_e-n_i$)/10$^8$ cm$^{-3}$ (b) and  
 potential profile along axis $r_p$ (normal to $B$-field) (c), 
 which starts from r=0, z=6 cm (solid line),
 6.5 cm (dashed line), 7 cm (dashed-dotted line),
 for B=100 G, $\alpha_B$=65$^{\circ}$ and $T_e$=10 eV.
}
\label{pot_q}
\end{figure}

 For the close $r_L$=0.075~cm, but for larger $T_e$=10~eV and B=100~G,
 the plasma parameters in Fig.\ref{pot_q} look similar to
 the case shown in Fig.\ref{pot_q100}. 
 The electron density has maxima and minima 1.3$\times$10$^{8}$cm$^{-3}$ 
 and 0.4$\times$10$^{8}$cm$^{-3}$, respectively. The negative and
 positive charge densities  are $\mp$0.25$\times$10$^{8}$cm$^{-3}$. 
  The potential profiles shown  in Fig.\ref{pot_q}(c) were taken across 
 the B-field vector starting from r=0 and z=6~cm,
 6.5~cm and 7~cm.  The double-layers across $B$-vector have 
 the potential drops of 0.2~V.  For these cases we can distinguish seven 
double-layers.

\section{Effect of Variation of Electron Temperature and B-Field }

\begin{figure}[h!]
\includegraphics[width=0.8\linewidth]{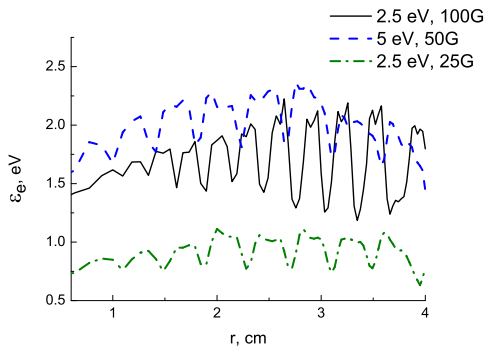}
\caption{
  Electron energy profile over radius for z=6 cm,
   $T_e$=2.5~eV, B=100~G (solid line),
   $T_e$=5~eV, B=50~G (dashed line),
   $T_e$=2.5~eV, B=25~G (dotted-dashed line),
    $\alpha_B$= 65$^{\circ}$.
}
\label{en}
\end{figure}

 As mentioned above the sheath with a small potential drop  
 $\delta \phi_w$=3~eV - 5~V 
 occurs over the grounded wall.
 The electrons with the energy larger than $\delta \phi_w$ 
 can overpass the sheath and escape from the plasma.
 Therefore the  temperature of plasma electrons  is smaller
 than $T_e$ which sets the Maxwellian velocity distribution for electrons
 during ion-electron pair generation. In simulations,
  the mean electron energy $\epsilon_e$ ranges from  1.6 eV to 3.4 eV for 
  $T_e$=2.5~eV - 10~eV for different strengths of B-field.
  An example of the plasma electron energy distribution for different
 $T_e$ and B is shown in Fig.\ref{en}. These energy profiles also
 reflect the specific structure of plasma in oblique magnetic confinement.
 As seen in Fig.\ref{en}, for the same $T_e$=2.5~eV
 the plasma electrons have essentially higher energy for B=100 G than
 for the case of B=25~G. 

\begin{figure}[h!]
\includegraphics[width=0.8\linewidth]{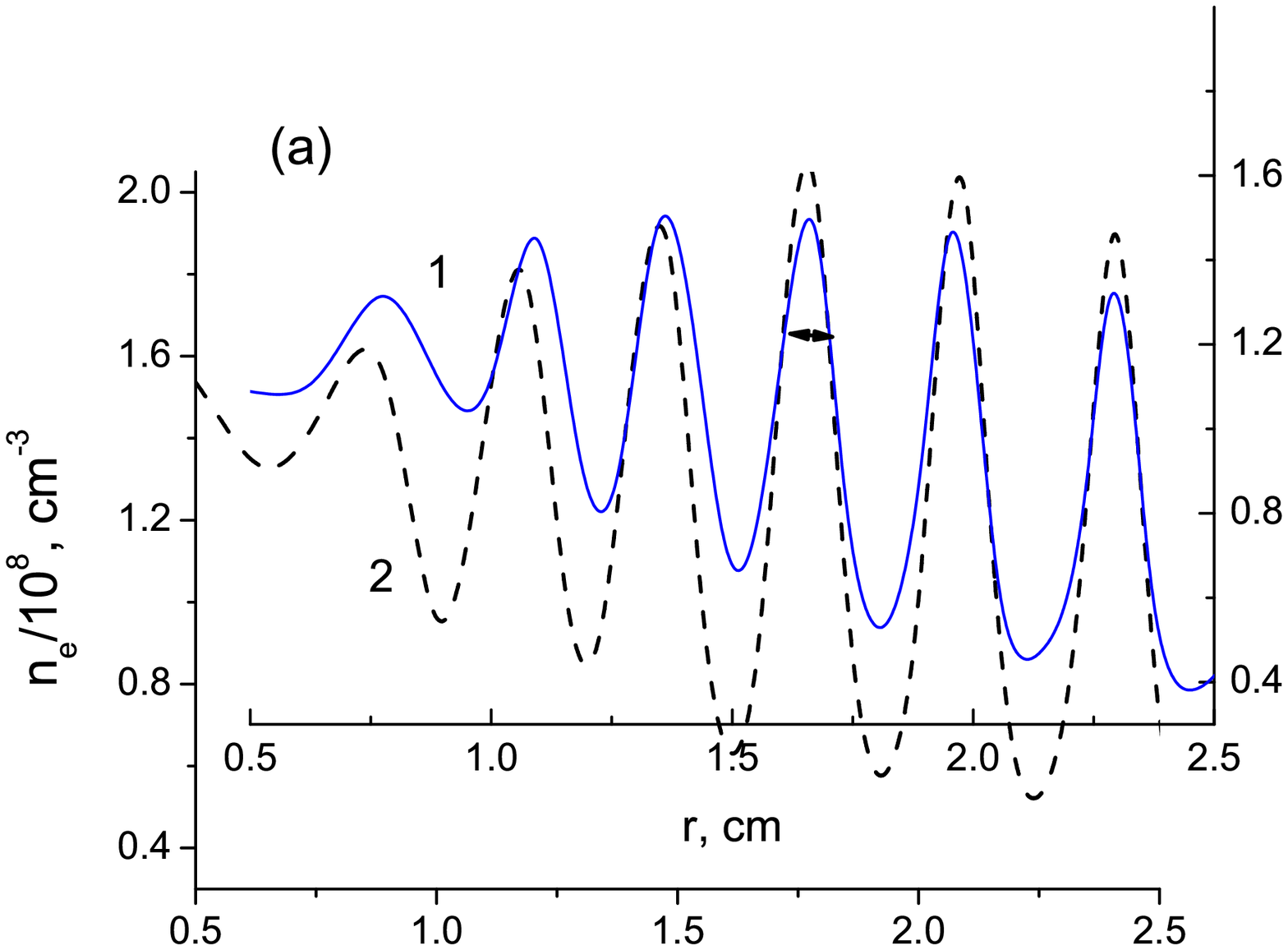}
\includegraphics[width=0.8\linewidth]{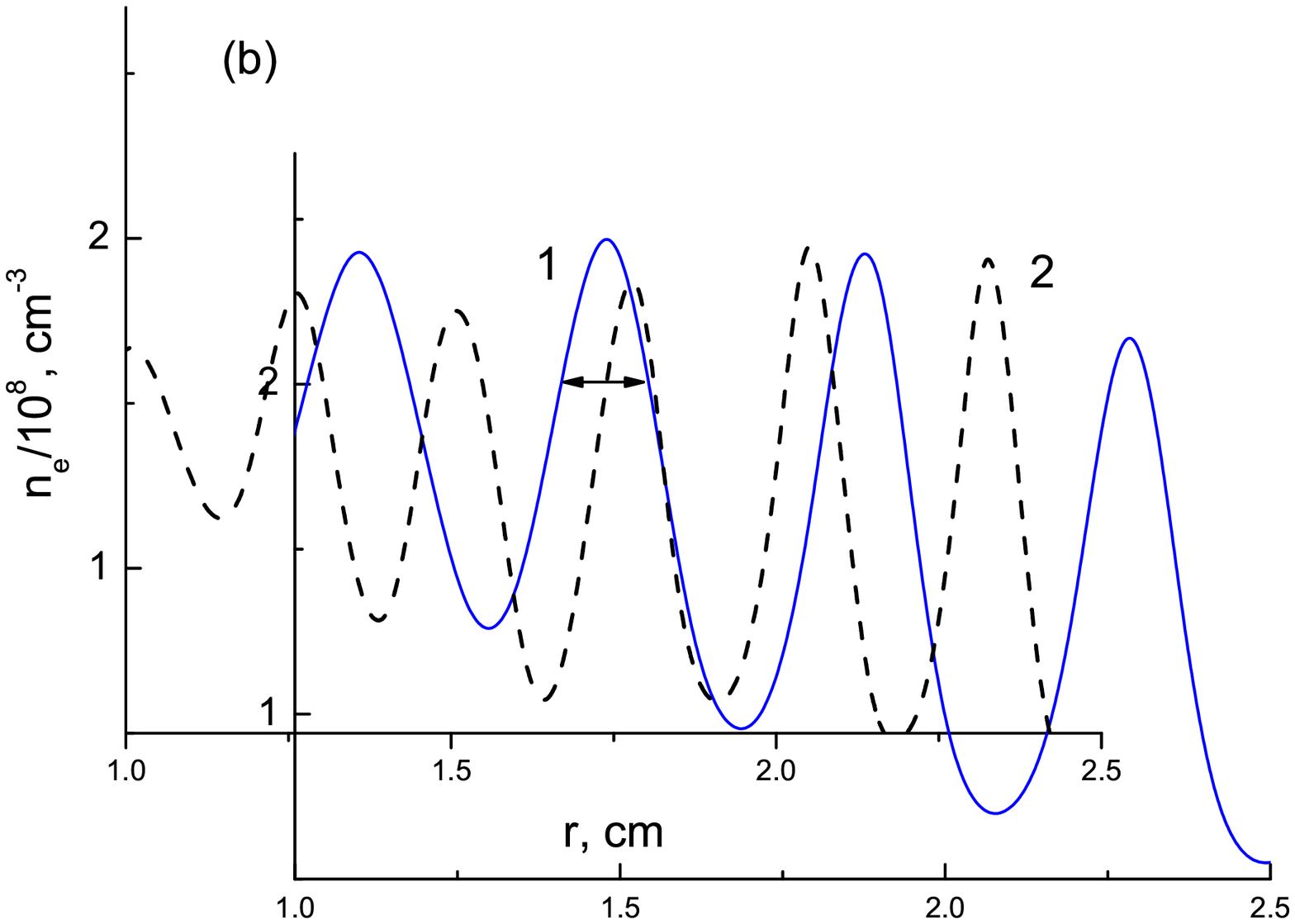}
\includegraphics[width=0.8\linewidth]{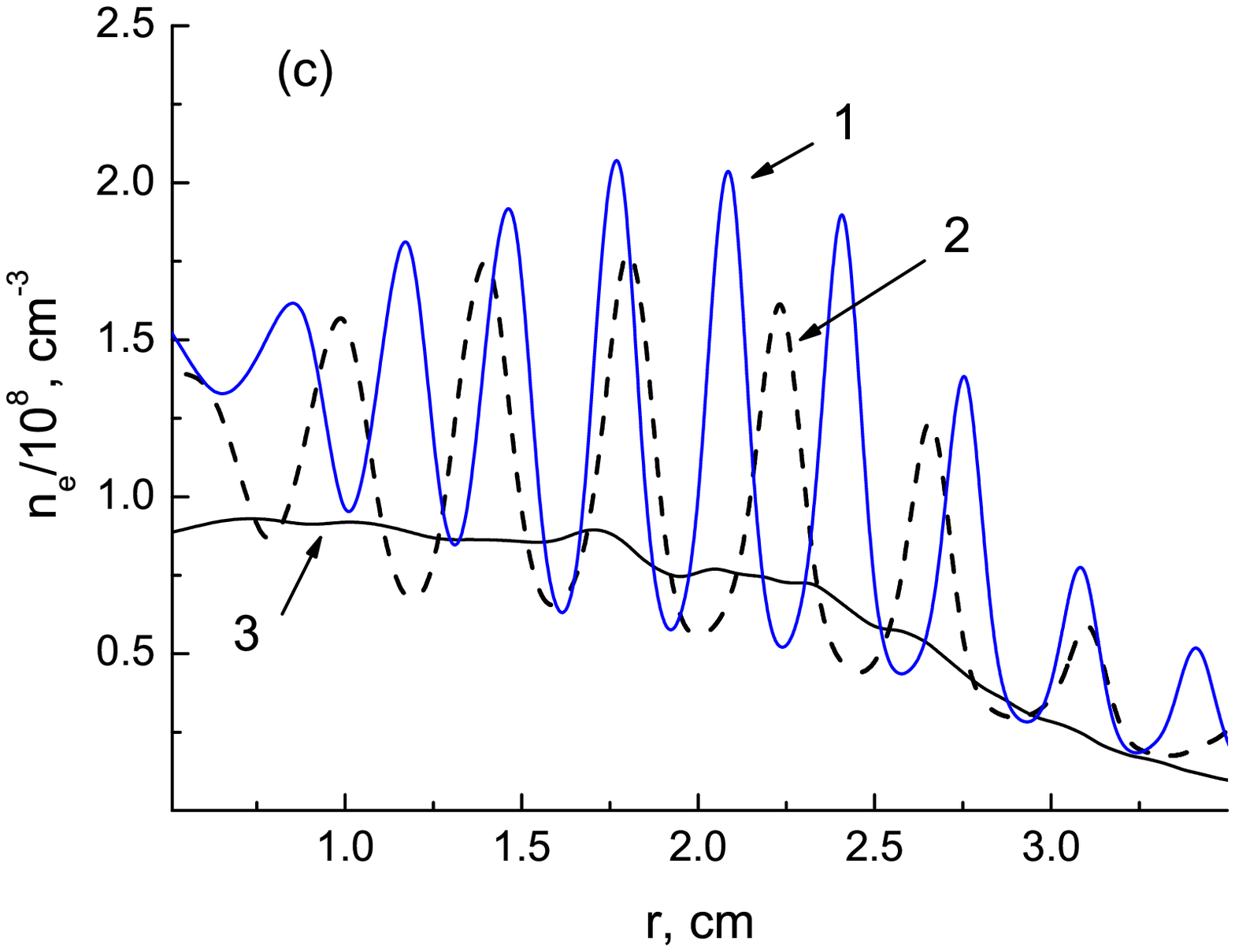}
\caption{
 Electron density profiles over radius at z= 6 cm 
 (a) for  $r_L$=0.075~cm ($T_e$=2.5~eV, B=50~G) (1) 
 and  $r_L$=0.05~cm ($T_e$=5~eV, B=100~G) (2),
 (b) for  $r_L$=0.15~cm ($T_e$=2.5~eV and B = 25~G) (1) 
and $r_L$=0.038~cm ($T_e$=2.5~eV and B = 100~G) (2),
 (c) for  B=50~G and $T_e$=2.5~eV ($r_L$=0.075~cm) (1), 5~eV $r_L$=0.105~cm (2) 
 and 10~eV $r_L$=0.15~cm (3).
  $\alpha_B$= 65$^{\circ}$. 
}
\label{larmor}
\end{figure} 
\begin{figure}[h!]
\includegraphics[width=0.8\linewidth]{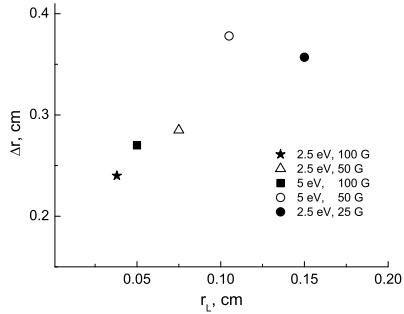}
\caption{ Period length of periodical structure from $r_L$ 
 for different $T_e$ and B,  $\alpha_B$= 65$^{\circ}$. 
}
\label{length}
\end{figure} 

 In the limit of large $r_L\sim T_e^{0.5}$/B and small $n_e$ the plasma 
 becomes smoother since
 the Larmor radius and inter-peaks distance are comparable. 
 With decreasing $T_e^{0.5}$/B and increasing $n_e$ the plasma forms
 more and more sharp peaks. We resolve 19 peaks for 
 the case of $\alpha_B$= 65$^{\circ}$ and $r_L$=0.07~cm  ($T_e$=2.5~eV
 and B=100~G).
 
 Let us consider a modification of plasma characteristics with changing
 parameter $r_L$ at given $\alpha_B$= 65$^{\circ}$. 
 The electron density profiles for two cases with close values of Larmor
 radii, 
 $r_L$=0.075~cm ($T_e$=2.5~eV, B=50~G) 
 and  $r_L$=0.05~cm ($T_e$=5~eV, B=100~G) 
 are shown in Fig.\ref{larmor}(a).
 The broadenings of these density peaks are similar and equal approximately 
 to 2$r_L$. 
 The $n_e$- profiles for two different values of $r_L$=0.15~cm
  ($T_e$=2.5~eV, B=25~G) and $r_L$=0.038~cm ($T_e$=2.5~eV, B=100~G) 
 are shown in Fig.\ref{larmor}(b).
 As expected the broadening of $n_e$-peaks is larger the former case
with larger $r_L$. 

 In Fig.\ref{larmor}(c), the $n_e$-profiles are shown for
 three values of $r_L$=0.075, 0.105, 0.15~cm.
 For these variants we took B=50 G and different $T_e$=2.5~eV, 
 5~eV and 10~eV. A smoothed profile refers to the case
 with $r_L$=0.15~cm and $T_e$=10~eV. 
  The $n_e$-profiles (3) Fig.\ref{larmor}(c) 
 and (1) in Fig.\ref{larmor}(b) have very different shapes whereas
  $r_L$=0.15~cm for both cases, but $\lambda_D$ is larger 
 for the former case. 
 
  With changing  $T_e$ from 5~eV to 2.5~eV,  
 the size of quasineutral area increases since 
 the wall potential drops from 5.1~eV to 3.6~eV.
 The number of ridges increases from 6 to 8 (see Fig.\ref{larmor}(c)),
 but inter-peak gap diminishes as the $\lambda_D$ decreases.
 It is seen that the characteristics of structure reflect of an interplay 
 of the wall potential drop, $r_L$ and $\lambda_D$.

  Period length of multi-steps double-layer structure is shown in
  Fig.\ref{length} as a function of $r_L$ for $\alpha_B$= 65$^{\circ}$. 
  Note that $\lambda_D$ slightly changes for different cases since 
 the plasma density depends on the potential drop near the wall,
 which in turn is a function of the electron energy.

 In conclusion, we have performed 2D3V PIC MCC simulations of dc
 discharge plasma in the cylindrical chamber at low gas pressure. The plasma
 is maintained by external ionization and confined by 
 external oblique magnetic field.
 The periodical structure with ridges of ion and electron densities have 
 been found for larger obliqueness of magnetic field. 
 The electron and ion ridges are shifted with respect to each other
 and double-layer structure appears across B-field and along the 
 potential rise. 

 The double-layers form due to a distortion of local quasineutrality
 in the presence of oblique magnetic field.
 When electron-ion pair appears after an ionization event  
 an electron begins Larmor gyramotion.  The electron is shifted from the ion
 in the direction normal to B-field and a local charge appears.

 The current flow channels are associated with ridges of 
 electron and ion densities. They are aligned with $B$-vector not 
 only in the area of quasineutral plasma,
 but also within the sheath over the wall.
 The modulation of electron current near the side wall in the 
 plasma under similar conditions
  was registered in Ref.\cite{disser}. 
 While our discharge geometry is not exactly the same we did not give the
 direct comparison of simulation and experimental results. However  
 the phenomena of multiple layer formation observed in our simulations and
 in the experiment \cite{disser} in discharge plasma induced by
 the oblique magnetic field are very similar. 

 The ion current approaching side wall of the cylindrical chamber 
 considerably increases and has a peaked profile in the case of 
 large obliqueness of B-field. The ion current distribution over the cathode
 sheath demonstrates some focusing effect in magnetic field 
 with larger angle $\alpha_B$. 

 The characteristics of plasma structure such as the number of peaks, 
 gap between them, their
 broadening depend on the Larmor radius ($\sim T_e^{0.5}/B$), 
 Debye length and the size of quasineutral plasma.
 The structure exists within some ranged of $T_e^{0.5}$/B and $n_e$. 
 With increasing  $T_e^{0.5}$/B and decreasing $n_e$
 the density peaks begin to overlap due to increasing broadening and 
 the plasma loses the periodical structure.

%%%%%%%%%%%%%%%%%%%%%%%%%%%%%%%%%%%%%%%%%%%%%%%%%%%%%%%%%%%%%%%%%%%%%%%%%%%%%%%%%%%%%%%%%%%%%%%%%%%%%%
\begin{acknowledgments}
The authors gratefully acknowledge FA9550-11-1-0160, 
Program Manager Mitat Birkan for support of this research.
One of the authors, IVS, was partly supported by grant of Russian
 Foundation of Basic Research No. 15-02-02536. 
\end{acknowledgments}

%%%%%%%%%%%%%%%%%%%%%%%%%%%%%%%%%%%%%%%%%%%%%%%%%%%%%%%%%%%%%%%%%%%%%%%%%%%%%%%%%%%%%%%%%%%%%%%%%%%%%%
\bibliographystyle{IEEEtran}

\end{document}